# ENERGY DENSITY FUNCTIONAL AND SENSITIVITY OF ENERGIES OF GIANT RESONANCES TO BULK NUCLEAR MATTER PROPERTIES

S. Shlomo[1] and A. I. Sanzhur[2]

[1]*Cyclotron Institute, Texas A&M University, College Station, USA*
[2]*Institute for Nuclear Research, Kyiv, Ukraine*

The development of a modern and more realistic nuclear energy density functional (EDF) for accurate predictions of properties of nuclei is the subject of enhanced activity, since it is very important for the study of properties of nuclear matter (NM), giant resonances and, in particular, of properties of rare nuclei with unusual neutron-to-proton ratios. Here, we provide a short review of the current status of the nuclear EDF and the theoretical results obtained for properties of nuclei and nuclear matter. We will first describe a method for determining the parameters of the EDF, associated with the Skyrme type effective interaction, by carrying out a Hartree-Fock based fit to extensive set of data of ground state properties and constraints. We will then describe the fully self-consistent Hartree-Fock based random-phase-approximation theory for calculating the strength functions $S(E)$ and centroid energies $E_{\mathrm{CEN}}$ of giant resonances and provide results for $E_{\mathrm{CEN}}$ of isoscalar and isovector giant resonances of multipolarities $L = 0 - 3$ for a wide range of spherical nuclei, using 33 EDFs associated with standard form of the Skyrme type interactions, commonly employed in the literature. We investigate the sensitivities $E_{\mathrm{CEN}}$ of the giant resonances to bulk properties of NM and determine constraints on NM properties, such as the incompressibility coefficient and effective mass, by comparing with experimental data on $E_{\mathrm{CEN}}$ of giant resonances.

## 1. Introduction

The atomic nucleus is a fascinating and important laboratory for the study of properties of a many body system with strongly interacting constituents. Energy density functional (EDF) theory provides a powerful approach for theoretical calculations of properties of many body systems. It is based on a theorem [1] for the existence of a universal energy density functional (EDF) that depends on the densities of the constituents and their derivatives, which leads to the exact value for the ground state energy by minimization procedure. However, the main challenge is to find the EDF. An important task of the nuclear physics community is to develop a modern EDF which accounts for the effects of few body and many body correlations and provides enhanced predictive power for properties of nuclei and nuclear matter (NM), such as the NM incompressibility coefficient, $K_{NM}$, and the density dependence of the symmetry energy, $E_{sym}$, needed for determining the equation of states (EOS) of symmetric and asymmetric NM. It is well-known that knowledge of the NM equation of state is very important in the study of properties of nuclei, heavy-ion collisions and astrophysical phenomena [2,3].

The phenomena of collective motions of strongly interacting nucleons in the many-body system of the atomic nucleus have been the subjects of experimental and theoretical investigations for many decades [4-7]. Of particular interest are the determination of properties of isoscalar (isospin $T=0$) and isovector ($T=1$) giant resonances of various multipolarities [8]. Over the years the strength function distributions, $S(E)$, and centroid energies, $E_{CEN}$, of isoscalar and isovector giant resonances have been found to be sensitive to physical quantities of nuclear matter (NM), such as the incompressibility coefficient $K_{NM}$, and the effective mass $m^*/m$. The resulting constraints on the values of the bulk properties of NM can be used to determine the next generation energy density functional (EDF) [9], with improved predictive power.

Since the earlier work of Brink and Vautherin [10], continuous efforts have been made to readjust the parameters of the Skyrme-type effective nucleon-nucleon (NN) interaction [11,12] to improve the theoretical prediction of properties of nuclei [13,14]. Many Skyrme type effective nucleon-nucleon interactions of different forms were obtained by fitting the HF results to selected sets of experimental data [13,14]. We emphasize that here we consider a specific standard form of the Skyrme type interaction, with ten (10) parameters [15]. We note that for fixed values for the nuclear matter properties the corresponding values for the Skyrme parameters can be determine by using the relations between the bulk properties of symmetric nuclear matter and the Skyrme parameters [9]. However, this is not possible due to the experimental uncertainties in the values of the nuclear matter properties. It is common to determine the parameters of the Skyrme interaction by fitting experimental data on bulk properties of nuclei, such as binding energies and charge radii, and include the experimental data on nuclear matter properties as constraints. It is very important to note that in determining the parameters of the Skyrme interaction, various approximations, concerning: (i) The values of the neutron and proton masses; (ii) The spin-density terms may be ignored; (iii) The Coulomb exchange term is approximated or ignored; (iv) The center of mass correction to the energy is approximated. These approximations should be taken into account for a proper application of the specific interaction. In the following we describe the method of determining the parameters of the very successful modern KDE0v1 Skyrme interaction [9] by a fit of the Hartree Fock results to extensive experimental data of ground state properties of wide range of nuclei, excitation energies of the isoscalar giant monopole and including constraints such as the Landau stability conditions for nuclear matter.

In the next section we provide a short review of the formalism including: (i) the standard form of the Skyrme type interaction with ten (10) parameters and the corresponding EDF with the HF method for calculating ground state properties of nuclei; (ii) the RPA approach for calculating strength functions $S(E)$ and the centroid energies $E_{CEN}$ of isoscalar and isovector giant resonances; (iii) the Folding-model (FM) distorted wave Born approximation (DWBA) for calculating excitation cross section for giant resonances, and; (iv) the EOS of symmetric and asymmetric NM in terms of bulk properties of NM. In section 3 we present: (i) results of calculations of determination of the parameters of the standard Skyrme interaction; (ii) pointing out the consequences of carrying out fully self-consistent HF-based RPA calculations of $S(E)$ and $E_{CEN}$ of giant resonances; (iii) pointing out the importance of carrying out microscopic calculations of excitation cross sections of giant resonances; (iv) demonstrate the importance of carrying out a proper comparison between relativistic and non-relativistic calculations of $E_{CEN}$, and; (v) present results of the calculated centroid energies, $E_{CEN}$, of isoscalar ($T=0$) and isovector ($T=1$) giant resonances of multipolarity $L = 0 - 3$ in $^{40,48}$Ca, $^{68}$Ni, $^{90}$Zr, $^{116}$Sn, $^{144}$Sm and $^{208}$Pb nuclei, within fully self-consistent spherical Hartree-Fock (HF)-based random phase approximation (RPA) theory, using 33 effective nucleon-nucleon Skyrme type interactions of the standard form. We also calculate the Pearson linear correlation coefficient to investigate the sensitivity of the $E_{CEN}$ of each giant resonance of specific isospin and multipolarity to each bulk property of NM, such as the incompressibility coefficient, $K_{NM}$, the effective mass $m^*/m$, the symmetry energy coefficients at $\rho_0$: $J = E_{sym}[\rho_0]$, and its first and second derivatives $L$ and $K_{sym}$, respectively, and $\kappa$, the enhancement coefficient of the energy weighted sum rule (EWSR) of the isovector giant dipole resonance (IVGDR). By comparing to experimental data, we determined constraints on the properties of NM. In section 4, we present our summary and conclusions.

## 2. Formalism

### 2.1. Skyrme energy density functional

We adopt the following standard form for the Skyrme type effective NN interaction [15]:

$$V_{12} = t_0(1 + x_0 P_{12}^\sigma)\delta(\mathbf{r}_1 - \mathbf{r}_2) + \frac{1}{2}t_1(1 + x_1 P_{12}^\sigma)[\overleftarrow{k}_{12}^2 \delta(\mathbf{r}_1 - \mathbf{r}_2) + \delta(\mathbf{r}_1 - \mathbf{r}_2)\vec{k}_{12}^2]$$

$$+ t_2(1 + x_2 P_{12}^\sigma)\overleftarrow{k}_{12}\delta(\mathbf{r}_1 - \mathbf{r}_2)\vec{k}_{12} + \frac{1}{6}t_3(1 + x_3 P_{12}^\sigma)\rho^\alpha\left(\frac{\vec{r}_1 + \vec{r}_2}{2}\right)\delta(\mathbf{r}_1 - \mathbf{r}_2) + \quad (1)$$

$$iW_0 \overleftarrow{k}_{12} \delta(\mathbf{r}_1 - \mathbf{r}_2)(\boldsymbol{\sigma}_1 + \boldsymbol{\sigma}_2) \times \vec{k}_{12}.$$

where $t_i$, $x_i$, $\alpha$, and $W_0$ are the ten (10) parameters of the interaction and $P_{12}^\sigma$ is the spin exchange operator, $\boldsymbol{\sigma}_i$ is the Pauli spin operator, $\vec{k}_{12} = -i(\vec{\nabla}_1 - \vec{\nabla}_2)/2$, and $\overleftarrow{k}_{12} = -i(\overleftarrow{\nabla}_1 - \overleftarrow{\nabla}_2)/2$. Here, the right and left arrows indicate that the momentum operators act on the right and on the left, respectively. The Skyrme energy-density functional $H(r)$, associated with the interaction of Eq. (1), is given by [15],

$$H = K + H_0 + H_3 + H_{eff} + H_{fin} + H_{so} + H_{sg} + H_{Coul}, \quad (2)$$

where $K = \frac{\hbar^2}{2m}\tau$ is the kinetic-energy term. For the Skyrme interaction of Eq. (2), we have

$$H_0 = \frac{1}{4}t_0\left[(2 + x_0)\rho^2 - (2x_0 + 1)(\rho_p^2 + \rho_n^2)\right], \quad (3)$$

$$H_3 = \frac{1}{24}t_3\rho^\alpha\left[(2 + x_3)\rho^2 - (2x_3 + 1)(\rho_p^2 + \rho_n^2)\right], \quad (4)$$

$$H_{eff} = \frac{1}{8}\left[t_1(2 + x_1) + t_2(2 + x_2)\right]\tau\rho + \frac{1}{8}\left[t_2(2x_2 + 1) - t_1(2x_1 + 1)\right](\tau_p\rho_p + \tau_n\rho_n), \quad (5)$$

$$H_{fin} = \frac{1}{32}\left[3t_1(2 + x_1) - t_2(2 + x_2)\right](\nabla\rho)^2 - \frac{1}{32}\left[3t_1(2x_1 + 1) + t_2(2x_2 + 1)\right]\left[(\nabla\rho_p)^2 + (\nabla\rho_n)^2\right], \quad (6)$$

$$H_{so} = \frac{W_0}{2}\left[\mathbf{J}\cdot\nabla\rho + x_w\left(\mathbf{J}_p\cdot\nabla\rho_p + \mathbf{J}_n\cdot\nabla\rho_n\right)\right], \quad (7)$$

$$H_{sg} = -\frac{1}{16}(t_1 x_1 + t_2 x_2)\mathbf{J}^2 + \frac{1}{16}(t_1 - t_2)\left[\mathbf{J}_p^2 + \mathbf{J}_n^2\right]. \quad (8)$$

Here, $H_0$ is the zero-range term, $H_3$ the density dependent term, $H_{eff}$ an effective-mass term, $H_{fin}$ a finite-range term, $H_{so}$ a spin-orbit term, $H_{sg}$ is a term that is due to tensor coupling with spin and gradient and $H_{Coul}$ is the contribution to the energy-density that is due to the Coulomb interaction. In Eqs. (3) – (8), $\rho = \rho_p + \rho_n$, $\tau = \tau_n + \tau_n$ and $\mathbf{J} = \mathbf{J}_n + \mathbf{J}_p$ are the particle number density, kinetic-energy density and spin-density, respectively, with $p$ and $n$ denoting the protons and neutrons, respectively. Note that the additional parameter $x_w$, introduced in Eq. (7), allows us to modify the isospin dependence of the spin-orbit term. We have used the value of $\hbar^2/2m = 20.734$ MeV fm$^2$ in determining the parameters of the Skyrme interaction KDE0v1. We would like to emphasize that we have included the contributions from the spin-density term as given by Eq. (8), which is ignored in many Skyrme HF calculations. Although the contributions from the Eq. (8) to the binding energy and charge radii are not very significant, they are very crucial for the calculation of the Landau parameter $G'_0$. The corresponding mean-field $V_{HF}$ and the total energy $E$ of the system are given by

$$V_{HF} = \frac{\delta H}{\delta \rho}, \quad E = \int H(\mathbf{r}) d\mathbf{r}, \tag{9}$$

where, the Skyrme energy-density functional $H(r)$, is given in Eq. (2).

In a spherical nucleus, the single-particle wave function can be written as a product of the radial function $R_n(r)$, the spherical harmonic function $Y_{jlm}(\mathbf{r},\sigma)$ and the isospin function $\chi_{m_\tau}(\tau)$,

$$\phi_n(\mathbf{r},\sigma,\tau) = \frac{R_n(r)}{r} Y_{jlm}(\mathbf{r},\sigma) \chi_{m_\tau}(\tau). \tag{10}$$

Assuming a closed-shell spherical nucleus, we use Eq. (10) to achieve the final form of the HF equations for spherical coordinates:

$$\frac{\hbar^2}{2m_\tau^*(r)} \left[ -R_n''(r) + \frac{l_n(l_n+1)}{r^2} R_n(r) \right] - \frac{d}{dr}\left( \frac{\hbar^2}{2m_\tau^*(r)} \right) R_n'(r)$$

$$+ \left[ U_\tau(r) + \frac{1}{r}\frac{d}{dr}\left( \frac{\hbar^2}{2m_\tau^*(r)} \right) + \frac{\left[ j_n(j_n+1) - l_n(l_n+1) - \frac{3}{4} \right]}{r} W_\tau(r) \right] R_n(r) = \varepsilon_n R_n(r). \tag{11}$$

where $m_\tau^*(r)$, $U_\tau(r)$ and $W_\tau(r)$ are the effective mass, the single particle potential and the spin orbit potential, respectively. They are given in terms of the Skyrme parameters and the nuclear densities. An initial guess is taken for the single-particle wave functions such as WS wave functions. The HF equations are then solved by iteration.

### 2.2. Self-consistent Hartree-Fock based random phase approximation

The response function $S(E)$ of the many-body system to an external field described by the single-particle operator, $F = \sum f(r_i)$, is given by [16]

$$S(E) = \sum_\nu \left| \langle 0|F|\nu \rangle \right|^2 \delta(E - E_\nu) = \frac{1}{\pi} \text{Im} \int_0^\infty dr \, dr' \, f(r) G(r,r',E) f(r') \tag{12}$$

where $G$ is the particle-hole Green function and the sum is over all RPA states $\nu$ of energy $E_\nu$. The transition density $\rho_t$ associated with the strength in the region $E \pm \Delta E$ is obtained from:

$$\rho_t(r,E) = \frac{\Delta E}{\sqrt{S(E)\Delta E}} \int_0^\infty f(r') \left[ \frac{1}{\pi} \text{Im} \, G(r',r,E) \right] dr'. \tag{13}$$

The RPA Green function is given by

$$G^{RPA} = G^{(0)}\left(1 - \frac{\delta V}{\delta \rho} G^{(0)}\right)^{-1}. \tag{14}$$

Here, $V$ is the Hartree-Fock potential, having a functional dependence on $\rho$, the density. The unperturbed Green function $G^{(0)}$ is given in terms of the Hartree-Fock Hamiltonian $H$, its occupied eigenstates $\phi_h$, and the corresponding eigen-energies $\varepsilon_h$, as

$$G^{(0)}(r_1, r_2, \omega) = -\sum_h \phi_h(r_1)\left(\frac{1}{H - \varepsilon_h - \omega} + \frac{1}{H - \varepsilon_h + \omega}\right)\phi_h(r_2) . \tag{15}$$

The sum in (15) is on the occupied states; $(H - E)^{-1}$ is the Hartree-Fock Green function for a single particle propagated from $r_2$ to $r_1$.

The electromagnetic single-particle scattering operator for the isoscalar ($T = 0$) excitation of multipolarity $L$ is given by [17],

$$F_L = \sum_i f(r_i) Y_{L0}(i) , \tag{16}$$

and the corresponding isovector ($T = 1$) single-particle scattering operator is given by,

$$F_L = \frac{Z}{A}\sum_n f(r_n) Y_{L0}(n) - \frac{N}{A}\sum_p f(r_p) Y_{L0}(p) . \tag{17}$$

The $S(E)$ of the different multipolarities is then determined by: $f(r) = r^2$, for the isoscalar and isovector monopole ($L = 0$) and quadrupole ($L = 2$), $f(r) = r^3$ for the octupole ($L = 3$), $f(r) = r$ for the isovector dipole ($T = 1, L = 1$), and lastly $f(r) = r^3 - (5/3)\langle r^2 \rangle r$ for the isoscalar dipole ($T = 0, L = 1$). We point out that for the isoscalar dipole we subtract the contribution from the spurious state [18,19]. We calculate the energy moments of the $S(E)$ using

$$m_k = \int_{E_1}^{E_2} E^k S(E) dE , \tag{18}$$

where $E_2 - E_1$ is the appropriate experimental excitation energy range. The centroid energies of the resonances are then obtained using:

$$E_{CEN} = m_1 / m_0 . \tag{19}$$

For $E_1 = 0$ and $E_2 = \infty$, the first energy moment, $m_1$, of the isoscalar operator $F_L$ may also be directly obtained from the HF ground state wave function:

$$m_1(L, T = 0) = \frac{1}{4\pi}\frac{\hbar^2}{2m}\int_0^\infty g_L(r) \rho(r) 4\pi r^2 dr , \tag{20}$$

thus leading to energy weighted sum rule (EWSR) [4]. In Eq. (20) $\rho(r)$ is the ground state density obtained from the HF ground state of the nucleus, while $g_L(r)$ depends on the multipolarity, $L$, and its $f(r)$:

$$g_L(r) = \left(\frac{df}{dr}\right)^2 + L(L+1)\left(\frac{f}{r}\right)^2. \tag{21}$$

The isovector EWSR is related to Eq. (20) by:

$$m_1(L, T=1) = \frac{NZ}{A^2} m_1(L, T=0)[1 + \kappa - \kappa_{np}], \tag{22}$$

where $\kappa$ is an enhancement coefficient which is due to the momentum dependence of the effective nucleon-nucleon interaction, given for the standard Skyrme-type interaction Eq. (1) by:

$$\kappa = \frac{(1/2)[t_1(1 + x_1/2) + t_2(1 + x_2/2)]}{(\hbar^2/2m)(4NZ/A^2)} \frac{2\int g_L(r)\rho_p(r)\rho_n(r) 4\pi r^2 dr}{\int g_L(r)\rho(r) 4\pi r^2 dr}, \tag{23}$$

while the correction factor $\kappa_{np}$ arises from the small differences between the neutron and proton densities, or in other words because $\rho_n(r) - \rho_p(r) \neq \frac{N-Z}{A}\rho(r)$, and is obtained from:

$$\kappa_{np} = \frac{(N-Z)}{A} \frac{A}{NZ} \frac{\int g_L(r)[Z\rho_n(r) - N\rho_p(r)] 4\pi r^2 dr}{\int g_L(r)\rho(r) 4\pi r^2 dr}. \tag{24}$$

We note that here we adopt the methods of Refs. [16,17,20] in the numerical evaluation of the strength functions and centroid energies of the giant resonances.

2.3. DWBA calculations of excitation cross-section

The distorted wave Born approximation (DWBA) has been employed successfully for theoretical description of low-energy scattering reactions [21,22]. The DWBA differential cross section for the excitation of a nucleus by inelastic scattering by alpha ($\alpha$) particle, $\alpha + N \rightarrow \alpha + N^*$, is given by,

$$\frac{d\sigma^{DWBA}}{d\Omega} = \left(\frac{\mu}{2\pi\hbar^2}\right)^2 \frac{k_f}{k_i} |T_{fi}|^2, \tag{25}$$

where $k_i$ and $k_f$ are the initial and final linear momenta of the $\alpha$-nucleus relative motion, respectively, and $\mu$ is the reduced mass. The transition matrix element $T_{fi}$ is given by

$$T_{\mathrm{fi}} = \left\langle \chi_{\mathrm{f}}^{(-)} \Psi_{\mathrm{f}} \middle| V \middle| \chi_{\mathrm{i}}^{(+)} \Psi_{\mathrm{i}} \right\rangle , \qquad (26)$$

where $V$ is the α-nucleon interaction, $\chi_{\mathrm{i}}^{(+)}$ and $\chi_{\mathrm{f}}^{(-)}$ are the incoming and outgoing distorted wave functions of the relative α-nucleus motion, respectively, and $\Psi_{\mathrm{i}}$ and $\Psi_{\mathrm{f}}$ are the initial and final states of the nucleus, respectively. To calculate the transition matrix element $T_{\mathrm{fi}}$, Eq. (26), one can adopt the following approach. First, integrate over the coordinates of the nucleus (in $\Psi_{\mathrm{i}}$ and $\Psi_{\mathrm{f}}$) to obtain the transition potential

$$\delta U(\mathbf{r}) \sim \int \Psi_{\mathrm{f}}^{*} V \Psi_{\mathrm{i}} \, d\mathbf{r}_1 d\mathbf{r}_2 \cdots d\mathbf{r}_A \qquad (27)$$

as a function of the relative coordinate $\mathbf{r}$ between the projectile and the nucleus and then calculate the cross section using

$$\frac{d\sigma}{d\Omega} = \left(\frac{\mu}{2\pi\hbar^2}\right)^2 \frac{k_{\mathrm{f}}}{k_{\mathrm{i}}} \left| \left\langle \chi_{\mathrm{f}}^{(-)} \middle| \delta U \middle| \chi_{\mathrm{i}}^{(+)} \right\rangle \right|^2 . \qquad (28)$$

The cross section is calculated using a certain DWBA code with the transition potential $\delta U(r)$ and the optical potential $U(r)$ as input.

The folding model approach [21] to determine the optical potential $U(r)$ and transition potential $\delta U(r)$, as convolutions between the projectile-nucleon interaction $V(|\mathbf{r}-\mathbf{r}'|, \rho_0(r'))$ and the ground state and transition densities, respectively, is commonly used in theoretical descriptions of α-particle scattering [22]. The optical potential $U(r)$ is given by

$$U(r) = \int d\mathbf{r}' V(|\mathbf{r}-\mathbf{r}'|, \rho_0(r')) \rho_0(r') . \qquad (29)$$

Here, $\rho_0(r')$ is the ground state HF density of a spherical target nucleus and the α-nucleon interaction $V(|\mathbf{r}-\mathbf{r}'|, \rho_0(r'))$ is assumed to have the parameterized form,

$$V(|\mathbf{r}-\mathbf{r}'|, \rho_0(r')) = -V_0 \left(1 + \beta_V \rho_0^{2/3}(r')\right) e^{\frac{-|\mathbf{r}-\mathbf{r}'|}{\alpha_V}} - iW_0 \left(1 + \beta_W \rho_0^{2/3}(r')\right) e^{\frac{-|\mathbf{r}-\mathbf{r}'|}{\alpha_W}} \qquad (30)$$

Note that $V(|\mathbf{r}-\mathbf{r}'|, \rho_0(r'))$ is complex and density dependent [22]. The parameters $V_0$, $\beta_V$, $\alpha_V$, $W_0$, $\beta_W$, $\alpha_W$ in Eq. (30) are usually determined by a fit to the elastic scattering data. The transition potential, $\delta U_L(r, E)$, for an excited state with the multipolarity $L$ and excitation energy $E$, is obtained from:

$$\delta U_L(r,E) = \int d\mathbf{r}' \delta\rho_L(\mathbf{r}',E)\left[V(|\mathbf{r}-\mathbf{r}'|,\rho_0(r')) + \rho_0(r')\frac{\partial V(|\mathbf{r}-\mathbf{r}'|,\rho_0(r'))}{\partial\rho_0(r')}\right], \quad (31)$$

where $\delta\rho_L(\mathbf{r}',E)$ is the transition density for the excited state. We point out that within the "macroscopic" approach, commonly employed in experimental analysis of scattering data, one adopts a semi-classical collective model transition densities, $\rho_{coll}$, [19,21-24] with radial forms which are independent of the excitation energy and are derived from the ground state density using a collective model.

Another approach for evaluating the transition matrix element $T_{fi}$, usually employed in theoretical calculations, is to first integrate over the relative α-nucleus coordinates to obtain the scattering operator,

$$O \sim \int \chi_f^{(-)*} V \chi_i^{(+)} d\mathbf{r}, \quad (32)$$

and then calculate the matrix element $\langle\Psi_f|O|\Psi_i\rangle$ within a theoretical model for $\Psi_i$ and $\Psi_f$, using the HF ground state density in (29) and the HF-based RPA transition density in (31). Note that it is quite common in theoretical calculation to adopt for the operator $O$ in (32) the operators of Eqs. (16) and (17), for determining the strength function $S(E)$. Therefore, for a proper comparison between experimental and theoretical results for $S(E)$, one should adopt the "microscopic" folding model approach in the DWBA calculations of $\sigma(E)$, using HF ground state density in (29) and the HF-based RPA transition density in (31).

### 2.4. Equation of state of symmetric and asymmetric nuclear matter

In the vicinity of the saturation density $\rho_0$ of symmetric NM, the equation of state (EOS) can be approximated by

$$E_0[\rho] = E_0[\rho_0] + \frac{1}{18}K_{NM}\left(\frac{\rho-\rho_0}{\rho_0}\right)^2, \quad (33)$$

where $E_0[\rho]$ is the binding energy per nucleon and $K_{NM}$ is the incompressibility coefficient which is proportional to the curvature of the EOS, $K_{NM} = 9\rho_0^2 \left.\frac{\partial^2 E_0}{\partial\rho^2}\right|_{\rho_0}$. The EOS of asymmetric NM (ANM) can be approximated by

$$E[\rho_n,\rho_p] = E_0[\rho] + E_{sym}[\rho]\left(\frac{\rho_n - \rho_p}{\rho}\right)^2, \tag{34}$$

where $\rho_p$ is the proton density, $\rho_n$ is the neutron density and $E_{sym}[\rho]$ is the symmetry energy at matter density $\rho$, given by

$$E_{sym}[\rho] = J + \frac{1}{3}L\left(\frac{\rho - \rho_0}{\rho_0}\right) + \frac{1}{18}K_{sym}\left(\frac{\rho - \rho_0}{\rho_0}\right)^2, \tag{35}$$

where $J = E_{sym}[\rho_0]$ is the symmetry energy at saturation density $\rho_0$, $L = 3\rho_0 \left.\frac{\partial E_{sym}}{\partial \rho}\right|_{\rho_0}$, and $K_{sym} = 9\rho_0^2 \left.\frac{\partial^2 E_{sym}}{\partial \rho^2}\right|_{\rho_0}$. Therefore, to extend our knowledge of the EOS, accurate values of $K_{NM}$ and $E_{sym}[\rho_0]$ and its first and second derivatives are needed in the vicinity of the symmetric NM saturation density. Here we consider the sensitivity of the centroid energies of the isoscalar and isovector giant resonances to bulk properties of NM, such as $K_{NM}$, $E_{sym}$ and the effective mass $m^*/m$.

## 3. Results

### 3.1. Determination of the parameters of the Skyrme interaction

Many Skyrme type effective nucleon-nucleon interactions of different forms were obtained during the last five decades by fitting the HF results to selected sets of experimental data [9,14]. We emphasize that here we consider the specific standard form of Eq. (1) for the Skyrme type interaction. We note that for a fixed set of values for the bulk properties of nuclear matter (NM) the corresponding values for the Skyrme parameters can be determined by using the relations between the properties of nuclear matter and the Skyrme parameters [9,15]. However, this is not possible due to the large uncertainties in the experimental values of the bulk properties of NM. It is common to determine the parameters of the Skyrme interaction by fitting HF results to experimental data on properties of nuclei, such as binding energies and charge radii, and include the experimental data on bulk properties of NM as constraints. For example, in the case of the modern KDE0v1 Skyrme interaction [9] the parameters were determined by a fit of the Hartree-Fock results to experimental data for binding energies and charge radii of an extended set of ground states of nuclei, which include neutron rich as well as proton rich nuclei. The experimental data for the spin-orbit (S-O) splitting of the $2p$ neutrons and protons "bare" single particle orbits in the $^{56}$Ni nucleus and the rms radii for the $1d_{5/2}$, $r_\nu(\nu 1d_{5/2})$, and $1f_{7/2}$, $r_\nu(\nu 1f_{7/2})$, neutron orbits in $^{17}$O and $^{41}$Ca nuclei, respectively, were also included in the fit. We note, in particular, that the experimental data for the isoscalar giant monopole resonance (ISGMR) constraint energies $E_0$ for the $^{90}$Zr, $^{116}$Sn, $^{144}$Sm and $^{208}$Pb nuclei and the critical density $\rho_{cr}$, determined by imposing the Landau stability [25] conditions for nuclear matter, up to the value of $2.5\rho_0$ with an error of $0.5\rho_0$, where $\rho_0$ is the saturation density, were also included in the fit. Moreover, the values of the Skyrme parameters were constrained by the experimental data on the bulk properties of NM and by requiring that: (i) a positive slope for the symmetry energy density for $\rho < 3\rho_0$; (ii) a value of $\kappa = 0.1 - 0.5$ for the enhancement factor of the energy weighted sum rule for the isovector giant dipole resonance (IVGDR) and; (iii) a value of $G'_0 > 0$ for the Landau particle-hole interaction parameter at $\rho = \rho_0$. The simulated annealing approach was employed in the minimization procedure to determine the Skyrme parameters with the best fit to the experimental data (see Ref. [9]).

It is very important to note that in determining the parameters of the Skyrme interaction, various approximations were made in the literature concerning: (i) the values of the neutron and proton masses; (ii) the spin-density terms may be ignored; (iii) the Coulomb exchange term is approximated or ignored; (iv) the center of mass correction to the energy is approximated; (v) the contribution of charge dependence terms in the nucleon-nucleon interaction is usually neglected. These approximations should be taken into account for a proper application of the specific interaction. In Table 1 we present three parameter sets: for the SkM$^*$ force [26], which gives the realistic values of the nuclear matter incompressibility and the deformation energies of heavy nuclei, for the more recent Sly4 interaction [15] and the most recent KDE0v1 interaction [9]. We point out that knowledge of the surface energy of finite nuclei provides an additional relation involving Skyrme parameters $t_1$ and $t_2$. We note that in the mean-field, adjusted to reproduce the experimental data of charge root-mean-square (RMS) radii, the calculated Coulomb displacement energies of analog states are smaller than the experimental data by about 7% [27]. It was shown [27-30] that this discrepancy is due to the neglect of the contributions of charge dependence in the nuclear force and the effect of long-range correlations. A good approximation for accounting for these contributions and also obtaining a good fit for binding energies of proton rich nuclei is to eliminate the contribution of the exchange coulomb term from the Hamiltonian, i.e. taking $C_{ex} = 0$, as was done in determining the parameters of the KDE0v1 Skyrme interaction [9], see Table 1. We note that the Skyrme interaction KDE0v1 also reproduces the experimental data of neutron stars and fission barriers [13], which were not included in the fit for determining the parameters of KDE0v1.

Table 1. Parameters of the Skyrme interactions SKM$^*$ [26], Sly4 [15] and KDE0v1 [9] and some associated properties of symmetric nuclear matter. The values of $C_{ex}$ = 0 or 1 indicate whether the Coulomb exchange term is omitted or included in the Hamiltonian.

| Force | SkM$^*$ | Sly4 | KDE0v1 |
|---|---|---|---|
| $t_0$ (MeV fm$^3$) | -2645.0 | -2488.91 | -2553.0843 |
| $t_1$ (MeV fm$^5$) | 410.0 | 486.82 | 411.6963 |
| $t_2$ (MeV fm$^5$) | -135.0 | -546.39 | -419.8712 |
| $t_3$ (MeV fm$^{3(1+\alpha)}$) | 15595.0 | 13777.0 | 14603.6069 |
| $x_0$ | 0.09 | 0.834 | 0.6483 |
| $x_1$ | 0.0 | -0.344 | -0.3472 |
| $x_2$ | 0.0 | -1.0 | -0.9268 |
| $x_3$ | 0.0 | 1.354 | 0.9475 |
| $\alpha$ | 1/6 | 1/6 | 0.1673 |
| $W_0$ (MeV fm$^5$) | 130.0 | 123.0 | 124.4100 |
| $C_{ex}$ | 1 | 1 | 0 |
| $E_0[\rho_0]$ (MeV) | -15.78 | -15.97 | -16.23 |
| $K_{NM}$ | 216.7 | 229.90 | 227.54 |
| $\rho_0$ (fm$^{-3}$) | 0.16 | 0.16 | 0.165 |
| $m^*/m$ | 0.79 | 0.70 | 0.74 |
| $J$ (MeV) | 30.03 | 32.00 | 34.58 |
| $L$ (MeV) | 45.78 | 45.96 | 54.69 |
| $\kappa$ | 0.53 | 0.25 | 0.23 |

### 3.2. Consequences of violations of self-consistency in RPA calculation

Accurate experimental data on the strength distributions, energies and widths of various giant resonances exists for a wide range of nuclei [6,31]. At present, the results of our fully self-consistent and HF based RPA calculations for centroid energies of various giant resonances are accurate within $0.1-0.2$ MeV, comparable to the current experimental accuracy [31]. In the following we will describe and present results of the investigations leading to the resolutions of the longstanding discrepancy in the value of $K_{NM}$, as deduced using Skyrme interactions ($K_{NM} = 210$ MeV) and Gogny interaction ($K_{NM} = 230$ MeV).

Violation of self-consistency in the HF-based RPA calculations of properties of giant resonances, such as the response functions $S(E)$ and centroid energies $E_{CEN}$, are mainly due to the neglect of some components of the nucleon-nucleon interaction, such as the Coulomb and spin-orbit interaction, which were included in the HF calculations but not in the RPA calculations and to limitation in the configuration space (i.e. numerical accuracy). We point up that the fulfillment of the energy weighted sum rules (EWSR) of the giant resonances and obtaining the $L = 1, T = 0$ spurious state, associated with the center of mass motion, at zero energy are necessary conditions for self-consistency in the HF-based RPA calculations, but not sufficient. The effects of violations of self-consistency in HF-based RPA calculations of $S(E)$, $E_{CEN}$ and the transition densities $\rho_t$ of various giant resonances were investigated in detail [17-19,32]. Violations of self-consistency may have significant effects on the $S(E)$, $E_{CEN}$ and $\rho_t$ of the ISGMR and on the ISGDR [17,18].

The HF-based RPA results of the strength functions $S(E)$ of the ISGMR in $^{208}$Pb and $^{90}$Zr, obtained using the KDE0 Skyrme interaction [9], are shown in Fig. 1. The full line (SC) corresponds to the fully self-consistent calculations. The dashed line and the open circle line represent the results for $S(E)$ obtained by neglecting the spin-orbit and Coulomb particle-hole interactions in the RPA calculations, respectively. The results of similar calculations for isoscalar giant resonances of multipolarity $L = 0-3$, using the SGII Skyrme interaction [33], are shown in Fig. 2 for $^{100}$Sn. We point out the violations of self-consistency result in a reduction of about 1 MeV (of about a 7%) in the ISGMR energy of $^{208}$Pb. This shift leads to about a 14% decrease in the value of the nuclear matter incompressibility coefficient $K_{NM}$, which corresponds to a shift of 30 MeV in $K_{NM}$. Therefore, we conclude that when comparing with experimental data, the same value of $K_{NM}$ is obtained for different non-relativistic interactions, such as Skyrme and Gogny interactions, if the HF-based RPA calculations of the centroid energy of the ISGMR are fully self-consistent. As shown in Fig. 2, the effects of violations of self-consistency on $E_{CEN}$ of isoscalar giant resonances of multipolarity $L = 1-3$ are relatively small. Similar, relatively small shift in the values of the $E_{CEN}$ of the isovector giant resonances were obtain, see Ref. [17].

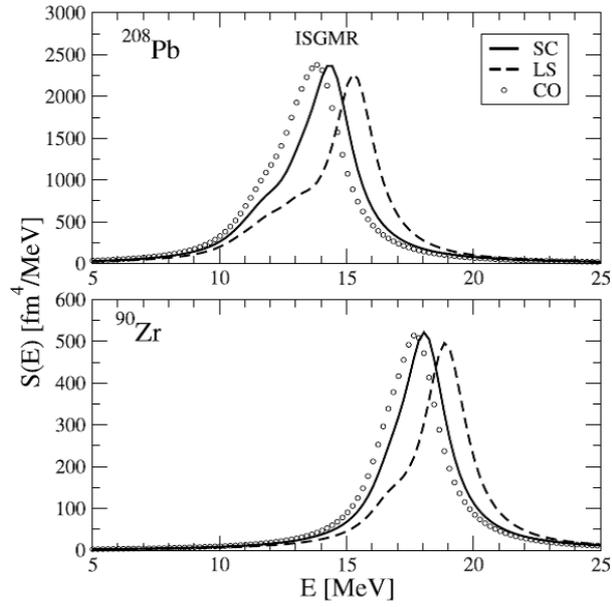

Fig. 1. Strength functions of isoscalar giant monopole for $^{208}$Pb and $^{90}$Zr nuclei calculated using the KDE0 interaction [9]. SC (full line) corresponds to the fully self-consistent calculation where LS (dashed line) and CO (open circle) represent the calculations without the particle-hole spin-orbit and Coulomb interactions in the RPA calculations, respectively.

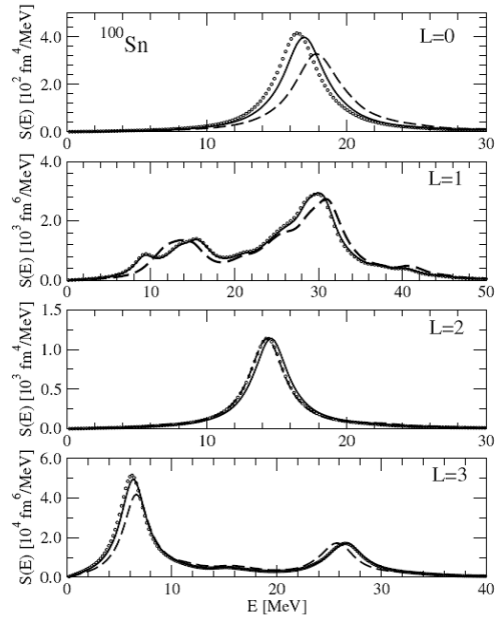

Fig. 2. HF-RPA results for the isoscalar strength functions of $^{100}$Sn for multipolarities $L = 0 - 3$ are displayed. SC (full line) corresponds to fully self-consistent calculations where LS (dashed line) and CO (open-circle) represent the calculations without the particle-hole spin-orbit and Coulomb interactions in the RPA calculations, respectively. The Skyrme interaction SGII [33] was used.

### 3.3. Nuclear matter incompressibility coefficient from the ISGDR

The isoscalar giant dipole resonance (ISGDR) is a compression mode and provides an independent source of information on the NM incompressibility coefficient $K_{NM}$. Early experimental investigations [34] resulted in a value of about 20 MeV for the $E_{CEN}$ of the ISGDR in $^{208}$Pb, which is smaller by about 4 MeV than the prediction of fully self-consistent HF-based RPA results obtained with interactions adjusted to reproduce experimental values of the $E_{CEN}$ for ISGMR in $^{208}$Pb. Therefore, the early experimental data on the $E_{CEN}$ of ISGDR leads to significantly smaller value of $K_{NM}$ (~ 170 MeV) than that obtained from the ISGMR, which raises some doubts concerning the unambiguous extraction of $K_{NM}$ from energies of compression modes of nuclei. To investigate this discrepancy we have therefore carried out microscopic calculation of the excitation cross section of the ISGDR, within the folding model (FM) distorted wave Born approximation (DWBA) using the HF ground state matter density and the RPA transition density, see section 2.3.

In Fig. 3 we present the results [19] of microscopic calculations of the excitation cross section $\sigma(E)$ of the ISGDR in $^{116}$Sn by 240 MeV α-particle scattering, carried out within the microscopic HF based RPA and the FM-DWBA theory [21,22] using the SL1 Skyrme interaction [35]. The solid line in the upper panel shows the HF-based RPA results for the fraction of the energy weighted sum rule, $ES(E)/EWSR$. The middle panel of the figure shows the double differential ISGDR cross sections at the angle of 1$^{st}$ maximum found using the transition potential for the ISGDR obtained from the HF-based RPA transition density. The lowest panel shows the results of the second panel (solid line) and from the collective model transition density $\rho_{coll}$ [36] (dashed line) both normalized to 100% of the EWSR of the ISGDR. Now, considering the solid line values of the cross section shown in the middle panel as the "experimental data" and dividing it by the cross-section values shown by the dashed line (semi-classical results) in the lower panel, we obtain the values of $ES(E)/EWSR$ shown by the dashed line in the upper panel, which is the result obtained in the experimental analysis of cross section data using the semi-classical form for the energy independent transition density.

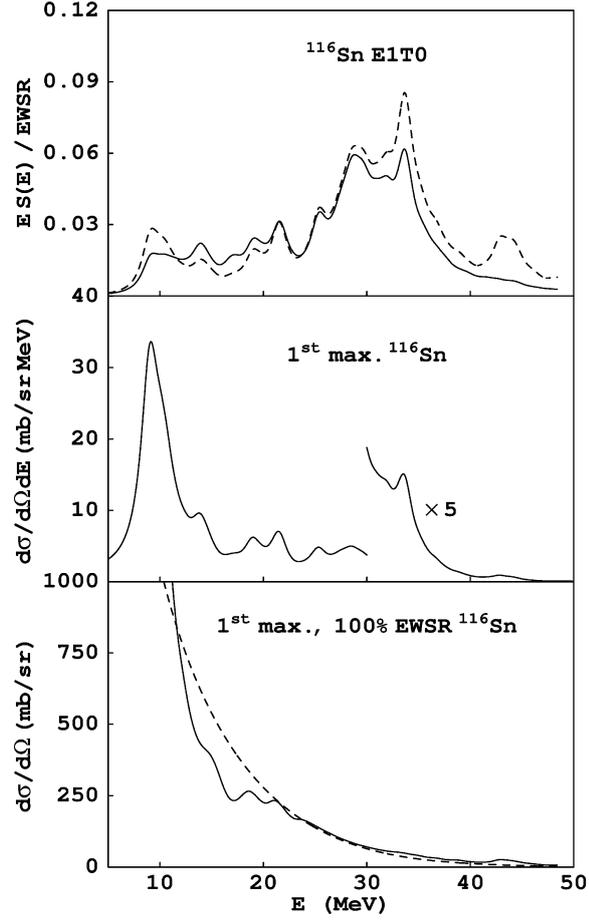

Fig.3. Reconstruction of the ISGDR EWSR in $^{116}$Sn from the inelastic α-particle cross sections. The middle panel: 1$^{st}$ maximum double differential cross section obtained from the RPA transition density $\rho_t$. The lower panel: maximum cross section obtained with the collective model transition density $\rho_{coll}$ (dashed line) and the HF-RPA transition density $\rho_t$ (solid line) normalized to 100% of the EWSR. Upper panel: The solid (dashed) line is the ratio between the middle panel curve and the solid (dashed) curve of the lower panel [19].

It is seen from the upper panel that using of the collective model transition densities $\rho_{coll}$ in analyzing the experimental cross sections increases the EWSR by about 15%. However, the shifts in the centroid energies are small (a few percent), similar in magnitude to the current experimental uncertainties. It is important to note [19] that the maximum cross section for the excitation of the ISGDR shown in the middle panel decreases strongly at high excitation energy and drop below the experimental sensitivity. This missing experimental strength leads to a reduction of about 3.0 MeV in the centroid energy of the ISGDR. Taking into account this missing strength significantly reduces the discrepancy between theory and experiment. This prediction was confirmed in an improved experiment [37]. Therefore, we conclude that the value of $K_{NM}$ deduced from the ISGDR compression mode is in good agreement with that deduced from the ISGMR compression mode.

### 3.4. Incompressibility coefficient of NM in relativistic and nonrelativistic models

Some relativistic RPA models yielded values of $K_{NM}$, deduced from the ISGMR, which are significantly larger than those obtained from the non-relativistic Skyrme HF-based RPA calculations. For example, the NL3 parameterization [38] is associated with $K_{NM} = 272$ MeV, as compared to the value of $K_{NM} = 240$ MeV deduced from the non-relativistic model. We have investigated this model dependence in Ref. [39] by generating parameter sets for Skyrme interactions by a least square fitting procedure using exactly the same experimental data for the bulk properties of nuclei considered in Ref. [38] for determining the NL3 parameterization of the effective Lagrangian used in the relativistic mean field (RMF) models. It is important to point out that the symmetry energy coefficient $J$ and charge rms radius of $^{208}$Pb were constrained to be very close to 37.4 MeV and 5.50 fm, respectively, as obtained with the NL3 interaction and $K_{NM}$ was fixed in the vicinity of the NL3 value $K_{NM} = 271.76$ MeV.

Table 2 present the results for $E_{CEN}$ of the ISGMR for several nuclei, obtained within fully self-consistent HF-based RPA (see Ref. [32]), using the KDE0 [9], SK255 [39], and SGII [33] Skyrme interactions, and the results obtained within the relativistic mean-field (RMF)-based RPA using the NL3 interaction [38]. We also compare with the experimental data of Refs. [37,40] calculated using the experimental excitation energy range $(\omega_1 - \omega_2)$. We point out that the results of Table 2 demonstrate that $K_{NM}$ can be deduced in a model independent way using relativistic and non-relativistic RPA calculations. Therefore, we conclude that $K_{NM} = 240 \pm 20$ MeV. The uncertainty of 20 MeV is mainly due to the uncertainty in $E_{sym}[\rho]$ and the possible effects of correlations beyond mean-field-based RPA. Note the difference in the value of $J$ associated with SGII and the SK255 Skyrme interactions, and with the NL3 interaction shown in Table 2.

Table 2. Results of fully self-consistent RPA calculations for the centroid energies of the ISGMR for interactions with various values of $K_{NM}$ and $J$ coefficients (in MeV)

| Nucleus | $\omega_1 - \omega_2$ | Experiment | NL3 | SK255 | SGII | KDE0 |
|---|---|---|---|---|---|---|
| $^{90}$Zr | 0-60 | | 18.7 | 18.90 | 17.89 | 18.03 |
| | 10-35 | 17.81±0.30 | | 18.85 | 17.87 | 17.98 |
| $^{116}$Sn | 0-60 | | 17.1 | 17.31 | 16.36 | 16.58 |
| | 10-35 | 15.85±0.20 | | 17.33 | 16.38 | 16.61 |
| $^{144}$Sm | 0-60 | | 16.1 | 16.21 | 15.26 | 15.46 |
| | 10-35 | 15.40±0.40 | | 16.19 | 15.22 | 15.44 |
| $^{208}$Pb | 0-60 | | 14.2 | 14.34 | 13.57 | 13.79 |
| | 10-35 | 13.96±0.20 | | 14.38 | 13.58 | 13.84 |
| $K_{NM}$ (MeV) | | | 272 | 255 | 215 | 229 |
| $J$ (MeV) | | | 37.4 | 37.4 | 26.8 | 33.0 |

### 3.5. Sensitivity of energies of giant resonances to properties of nuclear matter

The sensitivities of the strength function distributions, $S(E)$, and centroid energies, $E_{CEN}$, of isoscalar and isovector giant resonances of nuclei to the values of bulk properties of symmetric ($N = Z$) nuclear matter (NM): such as the binding energy per nucleon $E_0[\rho_0]$, the saturation density $\rho_0$, the incompressibility coefficient $K_{NM} = 9\rho_0^2 \left.\frac{\partial^2 E_0}{\partial \rho^2}\right|_{\rho_0}$, the symmetry energy coefficients at $\rho_0$, $J = E_{sym}[\rho_0]$, and its first and second derivatives $L = 3\rho_0 \left.\frac{\partial E_{sym}}{\partial \rho}\right|_{\rho_0}$ and $K_{sym} = 9\rho_0^2 \left.\frac{\partial^2 E_{sym}}{\partial \rho^2}\right|_{\rho_0}$, respectively, the effective mass $m^*/m$, and the enhancement coefficient $\kappa$ of the energy weighted sum rule (EWSR) of the isovector giant dipole resonance (IVGDR), have been investigated extensively very recently [8]. In this investigation:

1) The isoscalar and isovector giant resonances of multipolarities $L = 0$ to 3 were considered for the wide range of closed shell nuclei $^{40,48}$Ca, $^{68}$Ni, $^{90}$Zr, $^{116}$Sn, $^{144}$Sm and $^{208}$Pb. The occupation number approximation for the single-particle orbits for the open-shell nucleus $^{144}$Sm was adopted to insure a spherical nucleus.
2) Fully self-consistent Hartree-Fock (HF)-based random phase approximation (RPA) calculations of the centroid energies were carried out using 33 Skyrme effective nucleon-nucleon interactions of the standard form, Eq (1), commonly adopted in the literature. We note that wide ranges of values for the bulk NM properties are covered by the selected Skyrme interactions [8].

3) The Skyrme interactions were implemented in these calculations as they were designed. For example, by using the values of the masses of the proton and the neutron and the approximation for the Coulomb energy that were adopted in determining the parameters of the interactions. Self-consistency was ensured by including in the RPA calculations all the components of the interaction used in the HF calculation and carrying out highly accurate numerical calculations.

4) The sensitivity of $E_{CEN}$ to a NM property was deduced by calculating the corresponding Pearson linear correlation coefficient $C$, given, for quantities $x$ and $y$, by:

$$C = \frac{\sum_{i=1}^{n}(x_i - \bar{x})(y_i - \bar{y})}{\sqrt{\sum_{i=1}^{n}(x_i - \bar{x})^2}\sqrt{\sum_{i=1}^{n}(y_i - \bar{y})^2}}, \quad (36)$$

where $\bar{x}$ and $\bar{y}$ are the averages of $x$ and $y$ and the sum runs over all values. The different degrees of correlation can be classified as: strong ($|C| > 0.80$), medium ($|C| = 0.61 - 0.80$), weak ($|C| = 0.35 - 0.60$) and no correlation ($|C| < 0.35$).

In Table 3, the calculated Pearson linear correlation coefficients between different sets of NM properties are shown. We point out the weak correlation between $K_{NM}$ and $m^*/m$, the medium correlation between $m^*/m$ and the enhancement coefficient for the energy weighted sum rule (EWSR) of the IVGDR, $\kappa$, and the varying degrees of correlation between the symmetry energy coefficients $J$, $L$ and $K_{sym}$.

Table 3. Calculated Pearson linear correlation coefficients, $C$, for NM properties. The parameters of all 33 Skyrme effective nucleon-nucleon interactions were used to calculate $C$.

|  | $K_{NM}$ | $J$ | $L$ | $K_{sym}$ | $m^*/m$ | $\kappa$ | $W_0$ ($x_w = 1$) |
|---|---|---|---|---|---|---|---|
| $K_{NM}$ | 1.00 | 0.03 | 0.30 | 0.43 | -0.37 | -0.02 | 0.03 |
| $J$ | 0.03 | 1.00 | 0.72 | 0.49 | 0.07 | -0.24 | -0.25 |
| $L$ | 0.30 | 0.72 | 1.00 | 0.91 | -0.15 | -0.13 | -0.08 |
| $K_{sym}$ | 0.43 | 0.49 | 0.91 | 1.00 | -0.41 | -0.08 | 0.05 |
| $m^*/m$ | -0.37 | 0.07 | -0.15 | -0.41 | 1.00 | -0.63 | -0.19 |
| $\kappa$ | -0.02 | -0.24 | -0.13 | -0.08 | -0.63 | 1.00 | -0.03 |
| $W_0$ ($x_w = 1$) | 0.03 | -0.25 | -0.08 | 0.05 | -0.19 | -0.03 | 1.00 |

Table 4 presents the Pearson linear correlation coefficients between each nuclear bulk property of nuclear matter at saturation density and centroid energy of each giant resonance: the isoscalar giant monopole resonance (ISGMR), isoscalar giant dipole resonance (ISGDR), isoscalar giant quadrupole resonance (ISGQR), isoscalar giant octupole resonance (ISGOR), isovector giant monopole resonance (IVGMR), isovector giant dipole resonance (IVGDR), isovector giant quadrupole resonance (IVGQR) and isovector giant octupole resonance (IVGOR). Note that it is seen from Table 4 that, in particular, there exist strong correlations between $E_{CEN}$ and the incompressibility coefficient $K_{NM}$ for the ISGMR, between $E_{CEN}$ and the effective mass $m^*/m$ for the ISGQR and between $E_{CEN}$ and the enhancement coefficient $\kappa$ for the IVGDR energy weighted sum rule (EWSR) and, surprisingly, very weak correlations between $E_{CEN}$ and the symmetry energy $J$ or its first and second derivative, for the IVGDR.

Table 4. Pearson linear correlation coefficients between the centroid energy of each giant resonance and each nuclear matter property at saturation density.

|       | $K_{NM}$ | $J$   | $L$   | $K_{sym}$ | $m^*/m$ | $\kappa$ |
|-------|----------|-------|-------|-----------|---------|----------|
| ISGMR | 0.87     | -0.10 | 0.25  | 0.45      | -0.51   | 0.13     |
| ISGDR | 0.52     | -0.10 | 0.13  | 0.36      | -0.88   | 0.55     |
| ISGQR | 0.41     | -0.09 | 0.15  | 0.41      | -0.93   | 0.54     |
| ISGOR | 0.42     | -0.10 | 0.15  | 0.43      | -0.96   | 0.56     |
| IVGMR | 0.23     | -0.26 | -0.12 | 0.00      | -0.70   | 0.86     |
| IVGDR | 0.05     | -0.37 | -0.42 | -0.30     | -0.60   | 0.84     |
| IVGQR | 0.18     | -0.35 | -0.29 | -0.13     | -0.74   | 0.80     |
| IVGOR | 0.25     | -0.32 | -0.19 | 0.02      | -0.83   | 0.81     |

Figure 4 shows the $E_{CEN}$ of the ISGMR as a function $K_{NM}$ of the corresponding Skyrme interaction used in the calculation. Each nucleus is plotted separately, and the appropriate experimental band is contained by the dashed lines. Overall we see the well-known strong correlation between the $E_{CEN}$ and $K_{NM}$ [7,41], with a Pearson linear correlation coefficient $C \sim 0.87$ for all nuclei. It is interesting to note that we find a very weak correlation between $K_{NM}$ and the $E_{CEN}$ of the other compression modes, the ISGDR or the IVGMR (see Table 1). Figure 5 shows the $E_{CEN}$ of the ISGQR as a function of the effective mass $m^*/m$. We find a strong correlation between $E_{CEN}$ and $m^*/m$ (Pearson correlation coefficient $C = -0.93$). Figure 6 shows the $E_{CEN}$ of the IVGDR as a function of $\kappa$. We find a strong correlation between the values of the $E_{CEN}$ and $\kappa$ (Pearson correlation coefficient $C = 0.84$). As shown in Table 4 we find a very weak correlation between the values of the $E_{CEN}$ of the isovector giant dipole resonances we have and $J$ ($C = -0.37$), and similarly for its first derivative $L$ ($C = -0.42$) and second derivative $K_{sym}$ ($C = -0.30$). See Ref. [8] for other giant resonances.

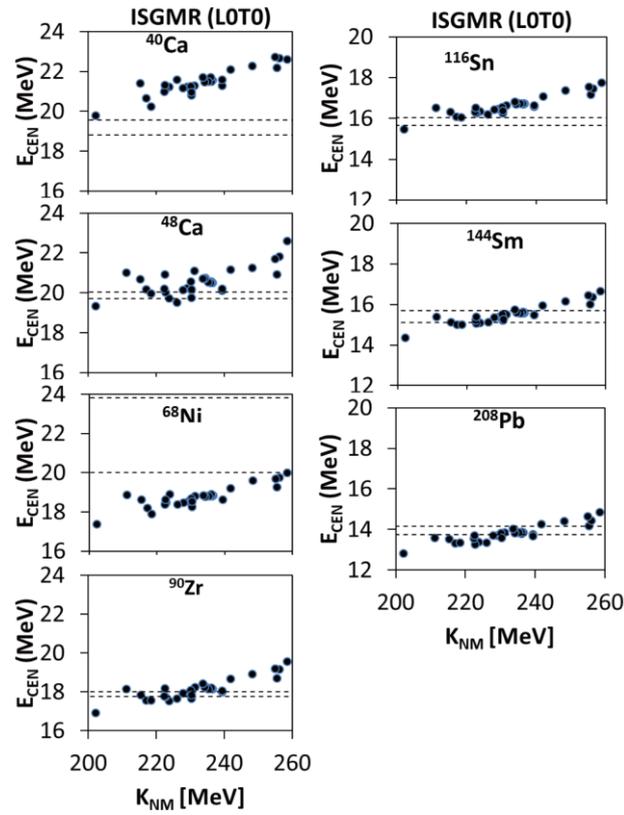

Fig. 4. Calculated centroid energies $E_{CEN}$ in MeV (full circle) of the isoscalar giant monopole resonances (ISGMR) for the different Skyrme interactions, as a function of the incompressibility coefficient $K_{NM}$. Each nucleus has its own panel and the experimental uncertainties are contained by the dashed lines.

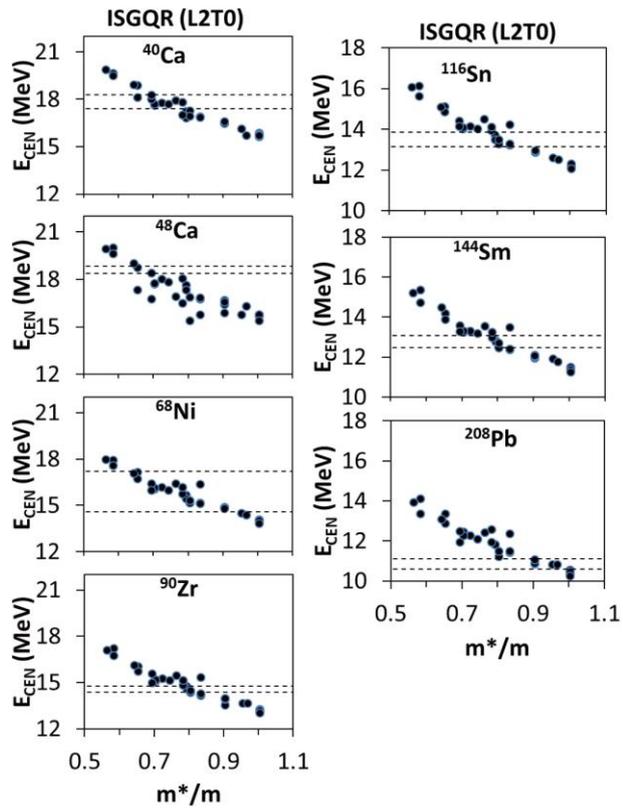

Fig.5. Similar to Fig. 4, for the isoscalar giant quadrupole resonance (ISGQR) as a function of the effective mass $m^*/m$.

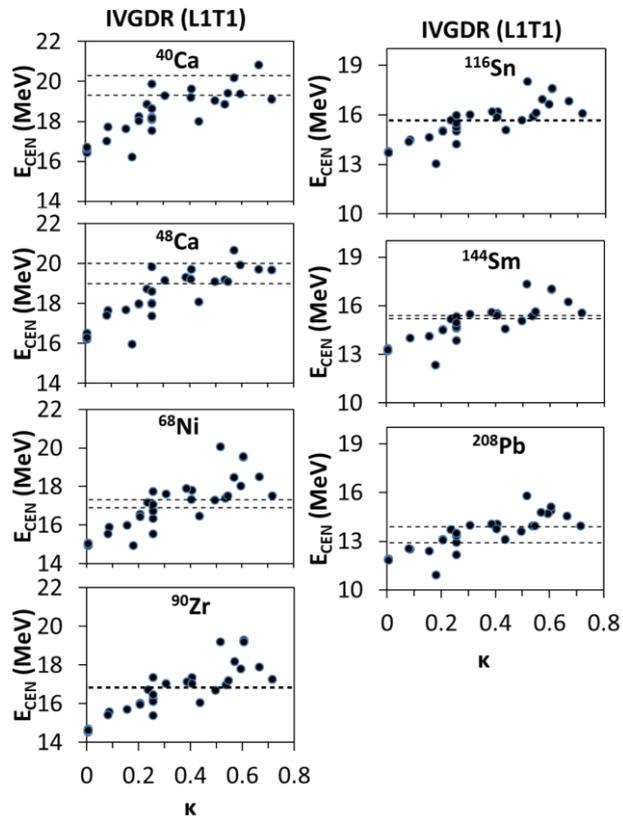

Fig.6. Similar to Fig. 4, for the isovector giant dipole resonance (IVGDR) as a function of the enhancement coefficient, $\kappa$, of the energy weighted sum rule (EWSR).

## 4. Summary and conclusions

We first described our knowledge of properties of isoscalar and isovector giant resonances of multipolarity $L = 0 - 3$, such as the strength functions $S(E)$ and centroid energies $E_{CEN}$ and their sensitivities to bulk properties of nuclear matter (NM). We then reviewed the current status of determining the parameters of modern energy density functional (EDF), associated with the standard form of the Skyrme type effective nucleon-nucleon interaction, having ten (10) parameters. In section 2 we have presented a short summary of the formalism of carrying out Hartree-Fock (HF) calculations of properties of ground states of nuclei, the formalism of carrying out HF-based random phase-approximation (RPA) calculations of $S(E)$ and $E_{CEN}$ and the folding model (FM) distorted wave Born approximation (DWBA) method for calculating the excitation cross section of giant resonance by inelastic scattering with a projectile such as the α particle.

The results of our calculations were presented and discussed in section 3. We first describe common approach employed in determining the parameters of the Skyrme interaction by a fit of HF results of properties of nuclei to experimental data, including constraints on NM properties. For example, the recent KDE0v1 was determined by a fit to binding energies and charge root-mean-square (rms) radii of nuclei ranging from the normal to exotic (proton- or neutron-rich) ones, and rms radii for $1d_{5/2}$ and $1f_{7/2}$ valence neutron orbits in the $^{17}$O and $^{41}$Ca nuclei, respectively. We have included in the fit the experimental data on the constraint energies of the isoscalar giant monopole resonance (ISGMR) and on the critical density $\rho_{cr} > 2\rho_0$, determined by the Landau parameters stability condition. Also included in the fit the constraints: (i) the slope of the symmetry energy must be positive for densities up to $3\rho_0$; (ii) the enhancement factor $\kappa$, of the energy weighted sum rule (EWSR) for the isovector giant dipole resonance (IVGDR), should lie in the range of 0.1 to 0.5; and (iii) the Landau parameter $G'_0$ of the particle-hole interaction, crucial for the spin properties of finite nuclei and nuclear matter, should be positive at $\rho = \rho_0$. We note that out of 240 Skyrme interactions, investigated by other researchers for the predictive power of these interactions, only the KDE0v1 also reproduce data on neutron stars and fission barriers that were not included in the fit. For comparison we presented in Table 1 the parameters and the associated bulk properties of NM of the popular Skyrme interactions, SkM$^*$, SLy4 and KDE0v1.

We also presented results of HF-based RPA calculations of $S(E)$ and $E_{\text{CEN}}$. We first considered problems of self-consistency in the calculations of $S(E)$ and $E_{\text{CEN}}$, result of microscopic calculation of excitation cross section of giant resonances needed for reliable determination of NM properties and demonstrated the model independence in deducing the incompressibility coefficient $K_{\text{NM}}$ from the ISGMR $E_{\text{CEN}}$, obtained using the HF-based RPA in non-relativistic or relativistic models approaches. We then presented results of $E_{\text{CEN}}$ and studied the sensitivities of $E_{\text{CEN}}$ to bulk properties of NM by employing 33 Skyrme type interactions, commonly used in the literature. We have demonstrated:

(i) the important effects of violation of self-consistencies (SC) in HF based RPA calculations of strength functions of giant resonances of multipolarities $L = 0 - 3$ and pointed out that due to a violation of SC the shift in the centroid energy of the ISGMR can be larger than 1 MeV (five times the experimental uncertainty), resolving the apparent dependence on the effective interactions in deducing $K_{\text{NM}}$ from the ISGMR;

(ii) by carrying out highly accurate microscopic calculations of excitation cross sections of the ISGMR and ISGDR and pointing out the missing strength at high excitation energy in experimental measurement of the of alpha excitation cross section of the ISGDR. This was confirmed by more accurate experiment, resolving the disagreement in the value of $K_{\text{NM}}$ deduced from the ISGMR or the ISGDR; and

(iii) by constructing Skyrme interactions with values of $K_{\text{NM}}$ similar to those obtained in the relativistic models we resolved the apparent model dependence in deducing the value of $K_{\text{NM}}$ from the $E_{\text{CEN}}$ of the ISGMR.

We have also presented results calculations of $E_{\text{CEN}}$, of the isoscalar ($T = 0$) and isovector ($T = 1$) giant resonances of multipolarities $L = 0 - 3$ in $^{40,48}$Ca, $^{68}$Ni, $^{90}$Zr, $^{116}$Sn, $^{144}$Sm and $^{208}$Pb, within the fully self-consistent spherical HF-based RPA theory, using 33 different Skyrme-type effective nucleon-nucleon interactions of the standard form commonly adopted in the literature. We reproduced the data for the $E_{\text{CEN}}$ of the ISGMR, ISGQR and IVGDR for most of the nuclei considered. For the ISGDR and ISGOR we found that most of the interactions are consistently higher than the experimental values for the centroid energy. We also studied the sensitivity of $E_{\text{CEN}}$ to bulk properties of nuclear matter (NM), such as the effective mass $m^*/m$, nuclear matter incompressibility coefficient $K_{NM}$, enhancement coefficient $\kappa$ of the energy weighted sum rule for the isovector giant dipole resonance and the symmetry energy $J$ and its first $L$ and second $K_{sym}$ derivatives at saturation density, associated with the Skyrme interactions used in the calculations. By comparing the calculated values of $E_{CEN}$ to the experimental data, we deduced constraints on the values of $K_{NM}$, $m^*/m$, and $\kappa$. We thus summarize our findings and conclude that:

- It is important to carry out fully self-consistent HF-based RPA calculations of $E_{CEN}$ to deduced model independent values for bulk properties of NM, in particular for the value of $K_{NM}$.
- It is important to carry out very sensitive measurement of the excitation cross section to deduce consistent values for bulk properties of NM from various giant resonances, particularly, for the $K_{NM}$ from the $E_{CEN}$ of the ISGMR and the ISGDR.
- We obtained strong, weak, and no correlations between the calculated values of $E_{CEN}$ and $K_{NM}$, for the compression modes of the ISGMR ($C \sim 0.87$), ISGDR ($C \sim 0.52$) and the IVGMR ($C \sim 0.23$), respectively.
- We obtained strong correlations between the effective mass $m^*/m$ and the calculated values of $E_{CEN}$ for the ISGDR ($C \sim -0.88$), ISGQR ($C \sim -0.93$), ISGOR ($C \sim -0.96$) and IVGOR ($C \sim -0.83$) and medium correlations for the IVGMR ($C \sim -0.70$), IVGDR ($C \sim -0.60$) and IVGQR ($C \sim -0.74$).
- We obtained strong correlations between the calculated values of the $E_{CEN}$ and the enhancement coefficient, $\kappa$, for the energy weighted sum rule of the IVGDR for all the isovector giant resonances considered ($C = 0.80 - 0.86$).
- We found weak to no correlations between the calculated values of $E_{CEN}$ and the symmetry energy coefficients $J$, $L$ or $K_{sym}$ for all the isovector resonances considered, see Table 4 for details.
- Considering the results of the $E_{CEN}$ of the ISGMR, ISGQR, and IVGDR of $^{40,48}$Ca, $^{68}$Ni, $^{90}$Zr, $^{116}$Sn, $^{144}$Sm and $^{208}$Pb we find that the interactions associated with NM properties in the following range best reproduce the experimental data: $K_{NM} = 210 - 240$ MeV, $m^*/m = 0.7 - 0.9$ and $\kappa = 0.25 - 0.70$.

We add that the constraints on NM properties that we obtained can be used to develop the next generation of energy density functionals by imposing the constraints in the fits used to determine the values of the parameters of the Skyrme interaction. We note that although these constraints may depend on the specific form of the interaction adopted, it is known that the centroid energy of the ISGMR is sensitive to $K_{NM}$. Similarly, the ISGQR is sensitive to the value of $m^*/m$ [42] because the effective mass influences the spacing between major nuclear shells and therefore the distribution of the response function. We also point out that when determining the best range for the effective mass we emphasized the results of the heavier nuclei more. Lastly, the dependence of the centroid energy of the IVGDR on $\kappa$, is expected from Eq. (22) for the energy weighted sum rule of the IVGDR, which is given by a constant times $(1+\kappa)$.

## Acknowledgements

This work is dedicated to our longtime friend and collaborator, Academician Professor Vladimir M. Kolomietz who died in June 2018, a personal loss to us and his family. S.S. is supported in part by the US Department of Energy, under Grant No DE-FG03-93ER40773. A.I.S. is partially supported by the Fundamental Research program "Fundamental research in high energy physics and nuclear physics (international collaboration)" at the Department of Nuclear Physics and Energy of the National Academy of Sciences of Ukraine.